# Numerical study of the influence of an applied electrical potential on the solidification of a binary metal alloy

P.A. Nikrityuk, K. Eckert, R. Grundmann
Institute for Aerospace Engineering
Dresden University of Technology,
D-01062 Dresden, Germany

## Abstract

In this work we study numerically the influence of a homogeneous electrical field on the fluid and heat transfer phenomena at macroscale and mesoscale during unidirectional solidification of a binary metal alloy. The numerical results showed that a pulse electric discharging applied perpendicularly to the solidification front leads to a much stronger Joule heating of the liquid phase in comparison to the solid phase. It was found that on the mesoscopic scale the electric current density is not homogeneous due to the complex shape of the dendrite and the difference between electrical conductivities of the solid and liquid phases. This inhomogeneity of the electrical current density in the mushy zone leads to the increase of the Joule heating of the dendrite in comparison to the interdendritic liquid and induces a pinch force (electromagnetic Lorentz force). The main features of the resulting convection in the interdendritic liquid are discussed.

## 1  Introduction

Control of solidification of metal alloys is one of the most demanding problems in the electromagnetic processing of materials. One of the innovative methods of such a control is the pulse electric discharging (PED). This method allows the modification of the microstructure during solidification [1-4]. The main feature of PED consists in a series of electric impulses passing through the solidifying melt. Due to the Joule heating caused by passing of an electric current, the temperature of the melt can increase. In the case of an inhomogeneous electrical current the interaction of the current with its own magnetic field produces a Lorentz force. This phenomenon, the so called pinch effect, received recently considerable attention in magnetohydrodynamics [5-8].

The pioneering work of the study of the influence of the direct electric current passing through the solidified melt were performed by Mirsa [1]. It was shown experimentally that the direct electric potential changes the nucleation and growth processes of the solid. But the mechanism of modification of the grains size were not understood. Nakada and coworkers [2] studied experimentally the influence of PED on the solidification structure of Sn15wt%Pb alloy. The electric discharging was carried out parallel to the solidification front by means of two cylindrical electrodes located along side wall of the cavity. The electric current was non-homogeneous. It was shown that solidification structures were modified from large grains with dendrites to finer grains with globular dendrites by means of pulse electric discharging with a capacitor bank. It was proposed that the Lorentz force (pinch force) induced at the moment of discharge is responsible for the break of dendrites into globular fragments due to high shear stress. But no numerical simulations were carried out to support this hypothesis.

To sum up, our understanding of the complex interaction between electrical current and solidified melt is far from being complete. On of the reasons for the still rather empirical applications of PED in unidirectional solidification of metals and alloys is the lack of detailed knowledge of the main mechanisms which are responsible for the grain size modification under the influence of electric current pulse. Motivated by this fact, this paper presents the first numerical study of the influence of PED on the heat and momentum transfer during directional solidification of Sn15wt%Pb alloy. In this paper we show the details about the development of the electro-vortex flow on mesoscale produced by interaction between the electric current passing through the melt and its own magnetic field. Furthermore we analyze the influence of the duration of the electric current pulse on the cooling curves.

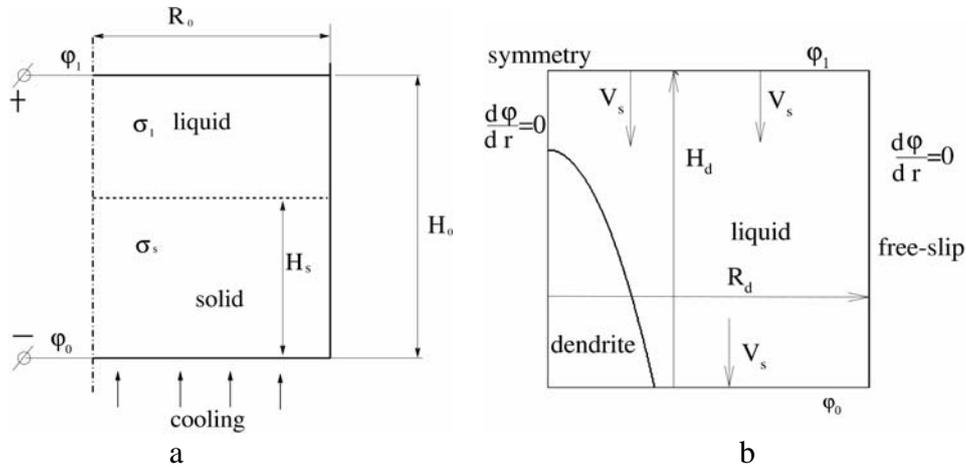

**Figure 1:** Schematic description of the geometry: axisymmetric cylindrical cavity on the macroscale (a) and columnar dendrite of parabolic shape on the mesoscale (b).

## 2    Problem formulation

To study the influence of the direct current applied during unidirectional solidification of a metal alloy on the macroscale heat transfer we consider a cylinder with non-conducting side walls of the height $H_0 = 0.075$ m and the radius $R_0 = 0.025$ m filled with the superheated alloy Sn15wt%Pb, see Fig. 1a. Between the bottom and the top of the cylinder an electric potential is applied. Thus a homogeneous electric current flows through the liquid and solid phases perpendicular to the solid front. Due to the homogeneity of the electric current there is no Lorentz force produced by the interaction of the current and its own magnetic field inside of the cavity on the macroscale [4]. The top and side walls of the cavity are thermally insulated, while the bottom is cooled at a rate governed by the instantaneous wall temperature, $T_W(t)$, and a uniform and constant overall heat transfer coefficient, $\alpha$:

$$q_W(t) = \alpha \cdot (T_W(t) - T_c), \tag{1}$$

where $T_c$ is the temperature of the cooling media. In this work the values for $\alpha$ and $T_c$ were set to 2100 W m$^{-2}$ K$^{-1}$ and 300 K, respectively to represent an intermediate velocity of solidification of about $V_s \approx 3 \cdot 10^{-4}$ m s$^{-1}$.

To study the physical processes in the mushy zone on the mesoscale we consider a columnar dendrite of a paraboloid shape without resolving the complex morphology of the its boundary, see Fig. 1b. The paraboloid shape is an accepted approximation of the dendrite, c.f. the famous work of Ivantsov [9]. The size of the domain considered is $R_d = 10^{-4}$ m which approximates typical dendrite arm spacing for $V_s \approx 3 \cdot 10^{-4}$ m s$^{-1}$ [10]. The center of cylindrical coordinate system lies on the axis of the symmetry of the dendrite, and is moving with the velocity corresponding to the solidification velocity $V_s$. Thus we have the dendrite which is flowed around the liquid phase which velocity equals to $V_s$, see Fig. 1b.

## 2.1 Electromagnetic field calculation

To calculate the electric current density we use Ohm's law:
$$\vec{j} = \sigma(\vec{E} + \vec{u} \times \vec{B}), \tag{2}$$

where $\sigma$ is the electrical conductivity of the mixture of solid and liquid phases, $\vec{u}$ is the velocity vector, $\vec{B}$ is the magnetic induction vector. By mesoscopic consideration of solidification $\sigma$ varies stepwise between solid and liquid phases. In the case of macroscopic consideration of the $\sigma$ variation a linear interpolation can be used:
$$\sigma = \sigma_l \varepsilon + \sigma_s (1 - \varepsilon), \tag{3}$$

where $\varepsilon$ is the volume fraction of liquid. In the liquid phase, corresponding to $\varepsilon = 1$, the electrical conductivity equals to $\sigma_l$. In the solid phase corresponding to $\varepsilon = 0$, the electrical conductivity equals to $\sigma_s$. The electric field intensity $\vec{E}$ is
$$\vec{E} = -\nabla \varphi. \tag{4}$$

Here $\varphi$ is the electric potential. To derive the electric potential we use the continuity condition of the electric current:
$$\nabla \cdot \vec{j} = 0. \tag{5}$$

Inserting eqs. (2) and (4) into eq. (5) written in cylindrical coordinates ($r, \theta, z$), we have:
$$\frac{1}{r}\frac{\partial}{\partial r}\left(r\sigma\frac{\partial \varphi}{\partial r}\right) + \frac{\partial}{\partial z}\left(\sigma\frac{\partial \varphi}{\partial z}\right) = -\frac{1}{r}\frac{\partial}{\partial r}(r\sigma u_z B_\theta) + \frac{\partial}{\partial z}(\sigma u_r B_\theta). \tag{6}$$

In this work we consider the axisymmetric case, thus $\dfrac{\partial \varphi}{\partial \theta} = 0$.

For the calculation of azimuthal magnetic field, $B_\theta$, we use Biot-Savart's law
$$\vec{j} = \frac{1}{\mu_0} \nabla \times \vec{B} \tag{7}$$

where $\mu_0 = 4\pi \cdot 10^{-7}$ H/m is the vacuum magnetic permeability. If there are no external magnetic fields we have only an azimuthal component of the magnetic induction, $B_\theta$, given by:

$$B_\theta = \frac{\mu_0}{r} \cdot \int_0^R r\, j_z\, dr \tag{8}$$

For the better understanding of the dynamics of the electrical filed parameters during solidification we simplify eq. (6) under the following conditions:
1. The electrical current density $\vec{j}$ is homogeneous.
2. During solidification there are only solid and liquid phases, i.e. no mushy zone exists.
3. The side walls of the cavity are isolated.

In this case eq. (6) has an analytical solution:

$$\varphi_{s-l} = \frac{\varphi_1 + \varphi_0 A_\sigma (A_s - 1)}{A_\sigma (A_s - 1) + 1} \tag{9}$$

where $\varphi_{s-l}$ is the electric potential on the boundary solid-liquid, $\varphi_0$ and $\varphi_1$ are the electric potentials on bottom and top, respectively, $A_s = H_0 / H_s$ and $A_\sigma = \sigma_s / \sigma_l$. Thus eq. (9) allows us to calculate electric field intensities, $E_s$, $E_l$ and Joule heating terms $\sigma_s E_s^2$, $\sigma_l E_l^2$ in solid and liquid phases, respectively. In this work we use eq. (9) for the validation of a solution of the eq. (6), see Section 3.

## 2.2 Macro-energy transport

In this study we restrict ourselves to the hypereutectic alloy Sn15wt%Pb which has the advantage of an initially stable stratification with respect to both the thermal and the solutal density change during unidirectional solidification. Thus without forced convection, the UDS of Sn15wt%Pb is not affected by thermosolutal convection. Furthermore shrinkage-driven flow is negligible. Since the homogeneous electric field does not induced convection the energy transport equation has the following form [10]:

$$\rho \frac{\partial}{\partial t}(c_p T) = \nabla \cdot (\lambda \nabla T) - \Delta H \rho \frac{\partial \varepsilon}{\partial t} + \sigma \vec{E}^2 \tag{10}$$

where $\lambda = \lambda_l \varepsilon + \lambda_s (1-\varepsilon)$ and $c_p = c_{p_l} \varepsilon + c_{p_s} (1-\varepsilon)$. The volume fraction of liquid $\varepsilon$ is calculated from the relation [11]:

$$\varepsilon = \left| \frac{T - T_s}{T_l - T_s} \right| \tag{12}$$

where $T_s = 183\,^\circ\text{C}$ is the solidus temperature, $T_l = 218.5\,^\circ\text{C}$ is the liquidus temperature. The last term in the eq. (10) is the Joule heat.

To further simplify the problem we evaluate the characteristic time scales. Namely the typical solidification time for $O(V_s) \approx 10^{-4}$ ms$^{-1}$ and $O(H_0) \approx 10^{-2}$ m is $O(10^2\, s)$. The characteristic time for solute diffusion on the macroscale is given by $O(\Delta z^2 / D_l) \approx 10^3$ s, where $\Delta z$ is the size of the control volume (CV) ($\Delta z = H_0 / 100$) and $D_l$ is the diffusion coefficient ($D_l = 1.5 \cdot 10^{-9}$ m$^2$ s). These different orders of magnitudes justify the neglect of solute mass transport on the macroscale.

**Table 1.** Physical properties of Sn15wt%Pb alloy

|  | Solid | Liquid |
| --- | --- | --- |
| Thermal conductivity $\lambda$, W m$^{-1}$ K$^{-1}$ | 57.99 | 26.2 |
| Specific heat $c_p$, J kg$^{-1}$ K$^{-1}$ | 210.85 | 233.8 |
| Molecular viscosity $\mu$, N s m$^{-2}$ | - | 1,873 10$^{-3}$ |

| Latent heat $\Delta H$, J kg$^{-1}$ | - | 54140 |
|---|---|---|
| Electric conductivity $\sigma$, A V$^{-1}$ m$^{-1}$ | 7.48 10$^6$ | 1.8 10$^6$ |

Material properties of Sn15wt%Pb alloy were calculated from a linear dependence on the mass concentration of its component and are given in the Tab 1. The material properties of pure Pb and Sn were taken from [12]-[14]. In this work we assume that the densities of the solid and liquid phases are identical and equal to 7889 kg m$^{-3}$.

## 2.3 Fluid flow on the mesoscale

On the microscale the electric current density is inhomogeneous due to the difference between $\sigma_s$ and $\sigma_l$, and complex form of the dendrites. The interaction between the electric current and its own magnetic field produces the pinch force. This force induces a forced convection in the mushy zone. To be able to capture the main flow structure taking place in the mushy zone we present a simplified model of the mesoscopic fluid flow. This model includes the Navier-Stokes equations decoupled from heat and mass transport in the liquid and solid phases. Here we assume that the dendrite is imbedded inside the heterogeneous fictious domain in which we globally solve the fluid dynamics problem. The corresponding N-S equations are based on the porous media theory, introducing the permeability relative to each phase [15]:

$$\nabla \cdot \vec{u} = 0 \tag{13}$$

$$\rho \frac{\partial \vec{u}}{\partial t} + (\rho \vec{u} \nabla) \vec{u} = -\nabla p + \mu \nabla^2 \vec{u} - \frac{\vec{u}}{K} + \vec{F}_L \tag{14}$$

where $K$ is permeability constant which prescribe immersed boundary conditions. This value related to each phase is defined by

$$K = \begin{cases} \infty, & \text{if } \varepsilon = 1 \\ 0, & \text{if } \varepsilon = 0 \end{cases} \tag{15}$$

The Lorentz force $\vec{F}_L$ have the following form:

$$\vec{F}_L = \vec{j} \times \vec{B} \tag{16}$$

Using eq. (2) the radial and axial projections of this force have the form:

$$F_{Lr} = \sigma \left( -E_z B_\theta - u_r B_\theta^2 \right) \tag{17}$$

$$F_{Lz} = \sigma \left( E_r B_\theta - u_z B_\theta^2 \right) \tag{18}$$

To justify the neglect of the heat and mass transport we compare the characteristic time scales for heat, mass and momentum transports with respect to the size of the control volume ($\Delta z = R_d / 100$). In particular the characteristic time for the Joule heat transfer in the liquid phase is $O(c_p \rho \Delta z^2 / \lambda) \approx 10^{-7}$ s. The characteristic time for the solute diffusion is given by $O(\Delta z^2 / D_l) \approx 10^{-3}$ s and the characteristic time for the viscous diffusion is $O(\rho \Delta z^2 / \mu) \approx 10^{-5}$ s. This evaluation shows that the mass transport is the slowest process, while temperature diffusion is the fastest one. Thus, a temperature perturbation produced by Joule heating is dissipated faster than fluid flow appears. Thus we assume that the liquid and the solid phases have the same temperature. In order to obtain first insights into the fluid flow induced by the pinch force we neglect the mass transport of the solute. However, to get the real insights into the transport process taking place on the mesoscopic scale during the PED it is necessary to consider momentum, energy and solute transports coupled with each other. This task is computationally demanding and will be done in future work.

# 3    Numerical scheme and code validation

The set of eqs. (6), (10), (13), (14) has been discretized by a finite-volume finite-difference based method. The time derivatives are discretized by a three-time-level scheme. The convection terms are discretized by a central difference second order scheme with deferred acorrection [15]. The system of linear equations is solved by using Stone's strongly implicit procedure (SIP). SIMPLE algorithm with collocated-variables arrangement was used to calculate the pressure and the velocities. Rhie and Chow stabilization scheme was used for the stabilization of pressure-velocity coupling. More details about the coupling algorithm can be found in [16].

Time marching with fixed time step was used. For every time step the outer iterations were stopped if residual of energy equation is less than $10^{-4}$ and less than $10^{-13}$ for pressure and momentum equations. Several grid-convergence and time-step-convergence tests were preformed to define proper grids and time steps leading to grid and time-step independent solutions. For the macro-scale energy transport simulations we used 20x70 grid, where first and second numbers correspond to the numbers of CV in the radial and axial directions, respectively, and a time step of 1 sec for diffusion controlled solidification with a PED duration of 30 sec. For the oscillating PED with a period $\Delta t_{off} = 1$ s between switch on and off, we used 20x200 grid with a time step of 0.2 s.

For the mesoscale simulations we used a structured non-uniform 350x350 grid to calculate the electric filed parameters and fluid flow for the case $V_s \ll 10^{-4}$ ms$^{-1}$. The time step used was set to $10^{-4}$ sec. For the calculation of fluid flow for the case $V_s = 10^{-4}$ ms$^{-1}$ we used a structured uniform grid 250x250 and the time step of $10^{-5}$ s.

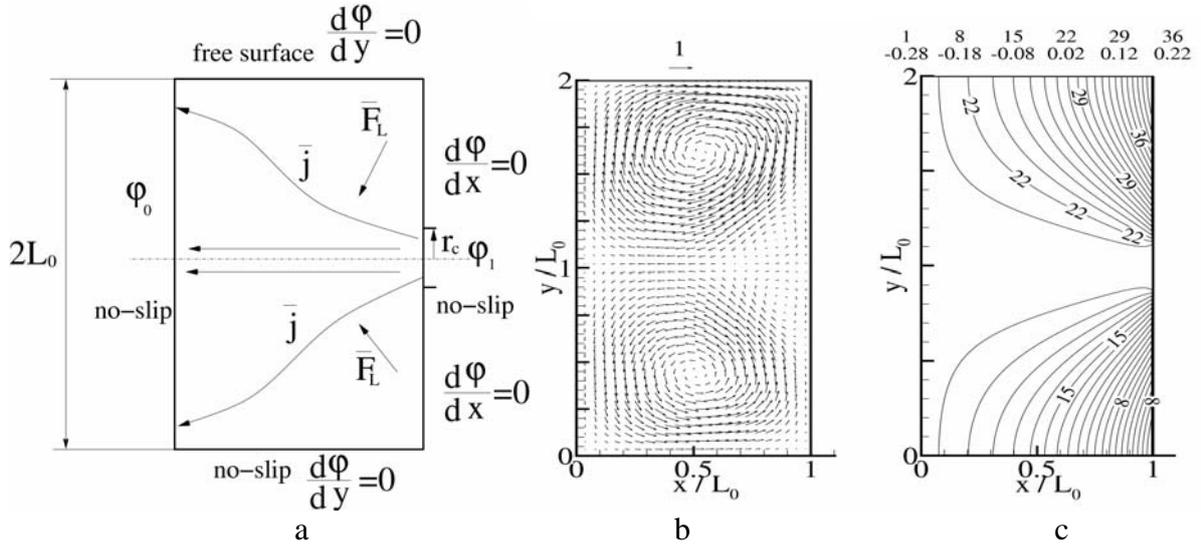

**Figure 2:** Code validations: scheme of the domain (a), vector plot of velocity scaled with $5 \cdot 10^{-4}$ m/s (b) and contour plot of magnetic induction $B_z$ scaled with $\mu_0 \sigma (\varphi_1 - \varphi_0) \cdot 0.5$ (c). Here we used $\varphi_0 = 0$ V, $\varphi_1 = 5 \cdot 10^{-4}$ V, $\sigma = 10^6$, $\rho = 6000$, $\mu = 2 \cdot 10^{-3}$, $L_0 = 35 \cdot 10^{-3}$ m, $r_c = 5 \cdot 10^{-3}$ m, 70x140 grid.

To validate the code we model the electro-vortex flow induced by an inhomogeneous direct electric current flowing between the sidewalls in a rectangular cavity. We consider the 2D case when the thickness of the cavity is much more less than both height and width. This ge-ometry is a simplified variant of the liquid metal current limiter (LMCL) investigated experimentally by Cramer et al. [7]. The 2D scheme of the device is shown in Fig. 2a. For the numerical simulations we used Cartesian coordinates ($x, y$). Fig 2b shows the spatial distribution of the velocity vectors induced by the Lorentz force. It can be seen that two large quasi symmetric vortices are generated. This is in good qualitatively agreement with the experiment [7]. The vortices induced are the product of the interaction between the electric current densities $j_x$, $j_y$ and its own magnetic field $B_z$ shown in Fig. 2c.

The validation of the solution of eq. (6) is done in the Section 4 by the comparison of the analytical solution (9) with the numerical one.

## 4 Results

The first series of numerical simulations is devoted to the study of the macro-energy transport during the solidification of the alloy under the influence of a pulsed electric current. Three cases were simulated. The first is the diffusion controlled UDS without PED, the second and third one concern the UDS with PED and voltages $\Delta\varphi = 0.05$ V and $\Delta\varphi = 0.1$ V, which was initialized after 50 sec of solidification and stopped after 80 sec of solidification. Fig. 3a shows the cooling curves obtained at the positions $z = 0.02$ m and $z = 0.065$ m from the bottom. It can be seen that during application of PED the cooling rate in the solid and liquid phases decreases. For $\Delta\varphi = 0.1$ the Joule heat is so large that the liquid phase is heated during the PED. To understand the impact of the PED on the spatial behaviour of the temperature we plot in the Fig. 4 the axial profiles of the temperature obtained at 70 sec. In comparison to the first case the temperature gradient on the boundary between the liquid and the mushy zone is increased. To investigate the influence of PED duration on the cooling curves we depict in the Fig. 3b the comparison of $T(t)$ gained for steady current and periodically switched on and off with the period $\Delta t_{off} = 1$ s. It is clearly seen that in the case of an oscillating PED the Joule heating effect decreases.

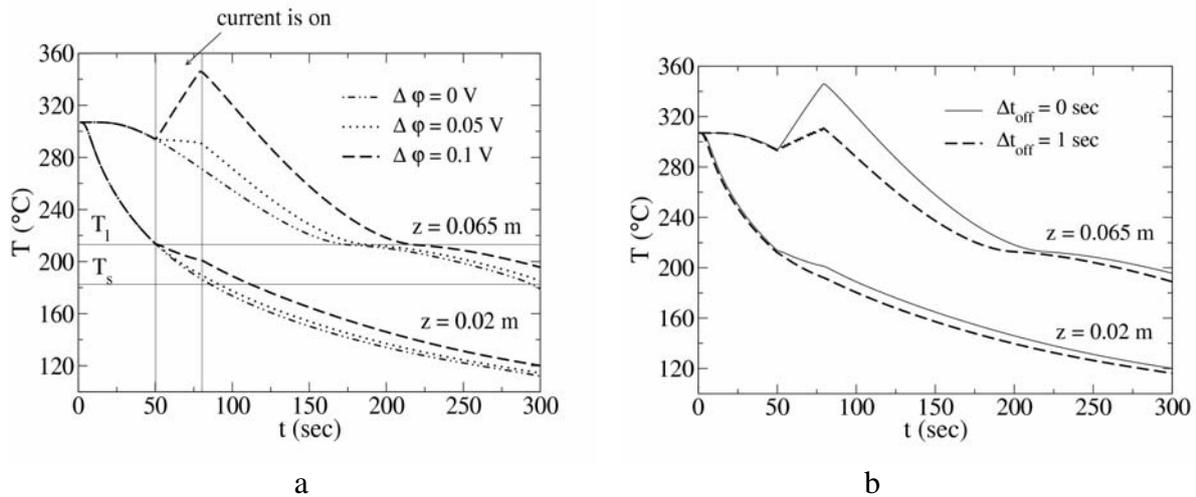

**Figure 3:** Predicted cooling curves: comparison of cooling curves gained for different voltage $\Delta\varphi$ (a) and comparison of cooling curves for $\Delta\varphi = 0.1$ V gained steady current and periodically switched on–off current (b). The period $\Delta t_{off}$ was set to 1 sec. Here $z$ is the distance from the chill.

To understand the increase of the temperature in the liquid phase we plot in the Fig. 5 the axial profiles of the nondimensional electric potential calculated by means of eq. (6) at $t = 70$ s. For comparison we depict the analytical solution for $\varphi$ in the case of a lacking mushy zone, i.e. the boundary between solid and liquid phases lies at $T_l$. For a better understanding of the Figure we plot additionally the profile of the liquid fraction $\varepsilon$ at that time. It can be seen that due to the higher electrical conductivity of the solid phase in comparison to the liquid we have a leap change in gradient of electric potential, in other words in $E_z$. Thus the electric field intensity in the solid phase is less than that in the liquid phase. As a result the Joule heating, $\sigma E_z^2$, in the liquid is higher than in the solid phase, see Fig. 6. This figure shows the dependency of the Joule heating on the ratio $H_s/H_0$, calculated by means of eq. (9).

The primary interest of the next series of calculations is the prediction of the evolution of the flow field induced by PED in the mushy zone and to study the distribution of electric potential, electric current and Joule heat on the mesoscopic scale, see Fig. 1b. For the calculation of electric parameters we used the following boundary conditions: Between the bottom and the top of domain we set the potential difference $\Delta\varphi$, taken from the macro-simulations, equal to $\Delta\varphi = 5\cdot10^{-4}$ V corresponding to a electric field intensity in the liquid phase of $E_z = -5\,\text{Vm}^{-1}$. Furthermore no current flows through the side walls in radial direction. Fig. 7 shows the predicted spatial distribution of the electric potential and electric current density vectors. It can be seen that due to the difference in electrical conductivities of the solid and the liquid phases the current density in the dendrite is higher than the current in the liquid. As a result the Joule heat increases in the dendrite in comparison to the liquid, see Fig. 8a. The Joule heat has the maximum value on the tip of the dendrite which is explained by maximum curvature of the surface in that place.

To study the fluid flow pattern induced by the pinch force we consider two cases. The first case corresponds to a very small solidification velocity $V_s \ll 10^{-4}\,\text{ms}^{-1}$. On referring to Sec. 2 this allows us to use the no-slip boundary conditions on the top, bottom and side wall of the domain, see Fig. 1b. The velocity vector plot, see Fig 8b, displays a toroidal vortex rotating in clockwise direction. Thus the interdendritic fluid flow washes the dendrite from the bottom to the tip, and in that way it will modify the solute boundary of the dendrite. As a result the shape of the dendrite will be changed. The second case is devoted to the growth of dendrite with an intermediate velocity $V_s = 1\cdot10^{-4}\,\text{ms}^{-1}$. Thus, a considerable relative flow around the dendrite occurs. In this case, the free-slip condition was used on the side wall, on the bottom: the velocity was set to $u_z = V_s$ and on the top outflow boundary condition was used. The velocity vectors plots calculated for PED with $E_z = -1\,\text{Vm}^{-1}$ and $E_z = -5\,\text{Vm}^{-1}$ are displayed in Fig. 9. It can be seen for the induced velocities smaller than the solidification velocity there is no change in the flow around the dendrite, see Fig. 9a. But if $u_z^{\max} > V_s$ a toroidal vortex rotating in clockwise direction appears near the tip of the dendrite, see Fig 9b. This vortex produces the upward flow which will transport the solute rejected by the dendrite to the upper part of the mushy zone. We suppose that this may lead to a larger constitutional undercooling.

To calculate the time required to establish a fully developed electro-vortical flow we introduce in analogy to [17] the volume-averaged meridional flow velocity as follows:

$$U_{rz} = \frac{2}{R_d^2 H_d} \int_0^{H_d} \int_0^{R_d} r\sqrt{u_r^2 + u_z^2}\, drdz \tag{19}$$

Its time history is presented in Fig. 10. The calculations were performed for two values of $E_z$. We found that the time of the flow establishment for all two $E_z$ is of order $O(10^{-3}\, s)$. This is a very promising result since the application of PED with a comparable pulse frequency would both avoid the Joule heating and reduce the consumptive power of the PED devise.

We argue that the fluid flow induced by the pinch force in the mushy zone can cause a mechanical fragmentation of dendrite. Probably the change of the solute concentration on the dendrite surface caused by the convection causes the constitutional fragmentation. Thus we suppose that the main mechanism of the grain refinement by the application of PED is related to the hydrodynamics of the turbulent regime which responsible for the refining of the grains by analog with works [17], [18],[19].

Summing up the results of simulations on macroscopic and mesoscopic scales, we are faced with a paradoxical situation. On the macroscale there is no convection due to the homogeneous electric current but on the mesoscale due the complex shape of the dendrite and $\sigma_s/\sigma_l \neq 1$, toroidal vortices appear in the interdendritic liquid which may induce macroscale convection. This assumption needs detailed consideration in the future work. We note that in the case of inhomogeneous electric current on the macroscale a pinch force appears on both macroscale and mesoscale.

## 5  Summary

The results of the numerical simulations of the heat transport on the macroscale showed that a pulse electric discharging applied perpendicularly to the solidification front leads to a much stronger heating of the liquid phase in comparison to the solid phase (the heating is caused by the Joule heating effect). We could show that a shorter duration of the PED pulses decrease the Joule heating of the melt. The numerical studies on the mesoscopic scale revealed that both due to the complex shape of dendrite and difference in electric conductivities between solid and liquid phases the Joule heating in the dendrite is increased in comparison to the heating in the interdendritic liquid. The Joule heating reaches its maximal value on the dendrite tip. The inhomogeneity of the electrical current density in the mushy zone induces a electromagnetic Lorentz force (pinch force). This force induces a toroidal vortex near the dendrite tip. It was shown that for the domain with side length of $10^{-4}$ m the time required for the flow to be established has the order of magnitude $10^{-3}$ s for $E_z$ of order of $O(10)$ Vm$^{-1}$.

## 6  Acknowledgements

The authors are grateful to Dr. M. Peric for the source code of the Navier-Stokes solver. We thank Armin Heinze for stimulating discussions. Financial support by the Deutsche Forschungsgemeinschaft (SFB609, B2) is gratefully acknowledged.

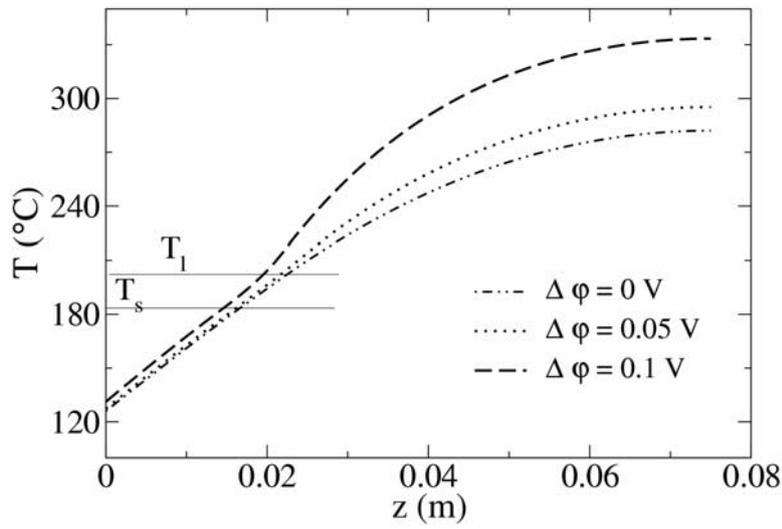

**Figure 4:** Predicted axial profiles of the temperature obtained for different voltages at $t = 70$ sec.

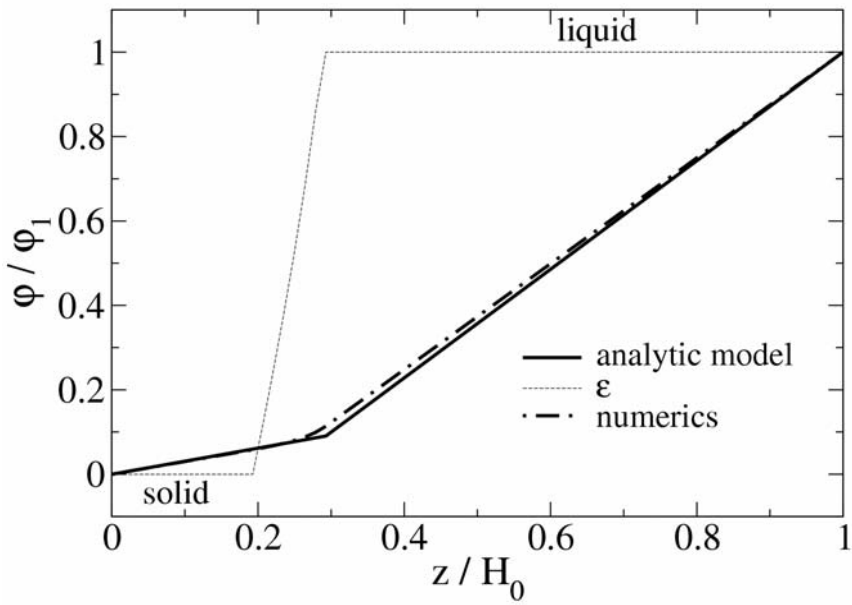

**Figure 5:** Comparison of the numerically and analytically calculated electric potential scaled with $\varphi_1$ at $t = 70$ sec. Here $\varepsilon$ is the volume fraction of the liquid. $\sigma_s / \sigma_l = 4.16$.

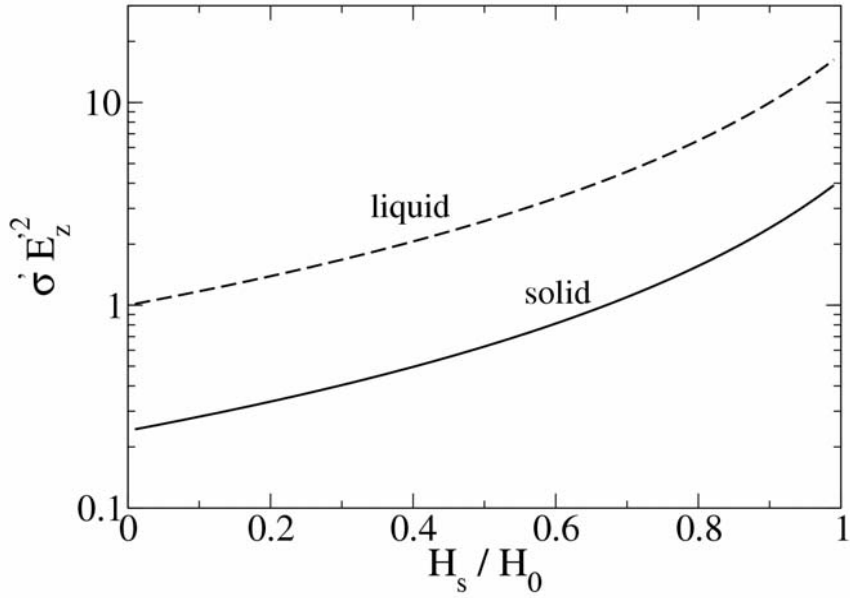

**Figure 6:** Analytically predicted dependence of the Joule heating term on the height of the solid phase $H_s$ scaled with the height of the cavity $H_0$. Here $\sigma' = \sigma_{l,s}/\sigma_l$, $E_z' = E_z H_0/\Delta\varphi$, $\sigma_s/\sigma_l = 4.16$.

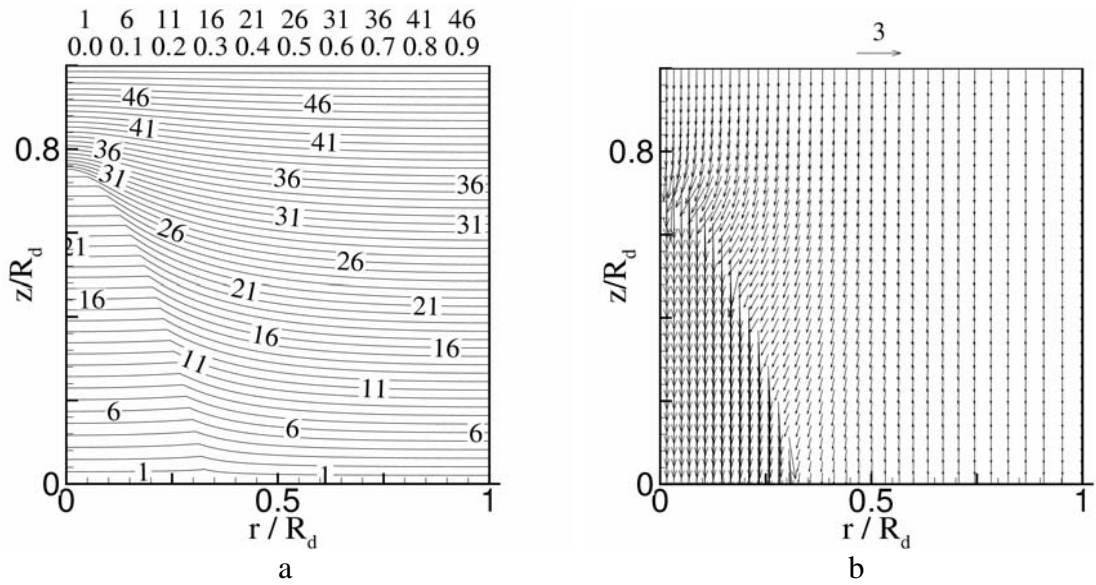

**Figure 7:** Predicted distribution of electric parameters: nondimensional electric potential $\varphi/\Delta\varphi$ (a) and vector plot of electric current density scaled with $\sigma_l \Delta\varphi/R_d$ (b). Here $R_d = 10^{-4}$ m, $\Delta\varphi = 5\cdot 10^{-4}$ V.

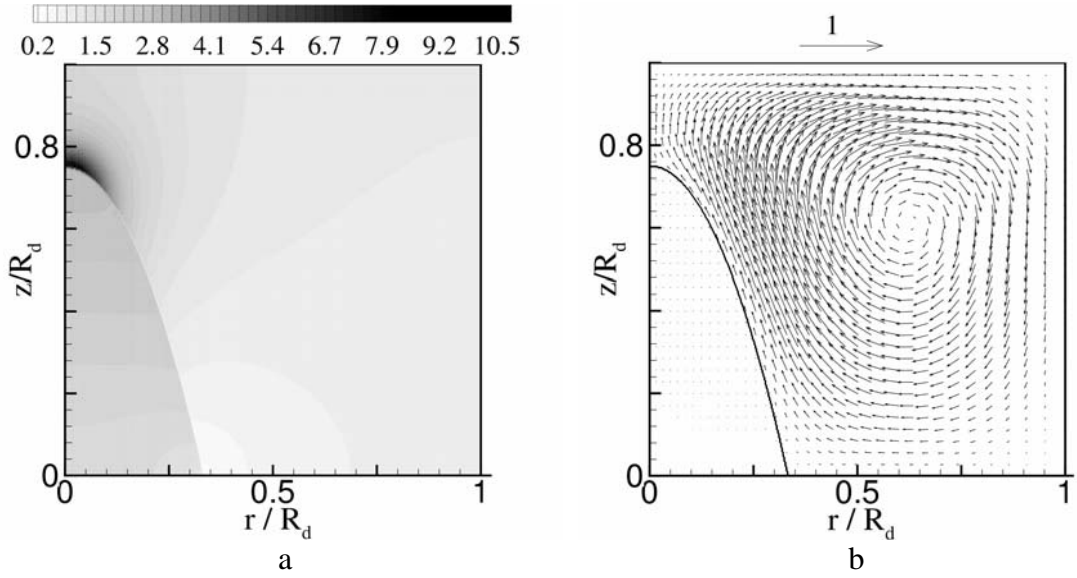

**Figure 8:** Predicted spatial distribution of the Joule heating term scaled with $\sigma_l (\Delta\varphi / R_d)^2$ (a) and meridional velocity scaled with $2.5 \cdot 10^{-4}$ ms$^{-1}$ (b). Here $\Delta\varphi = 5 \cdot 10^{-4}$ V.

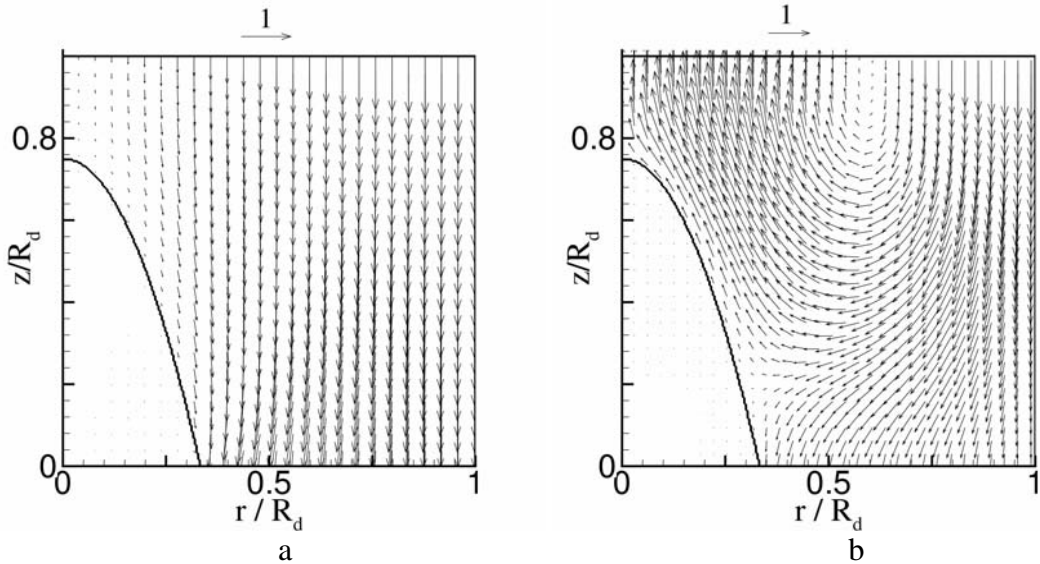

**Figure 9:** Predicted spatial distribution of meridional velocity scaled with $V_s = 10^{-4}$ m s$^{-1}$ for $E_z = -1$ Vm$^{-1}$ (a) and meridional velocity scaled with $2.5 \cdot 10^{-4}$ ms$^{-1}$ for $E_z = -5$ Vm$^{-1}$ (b)

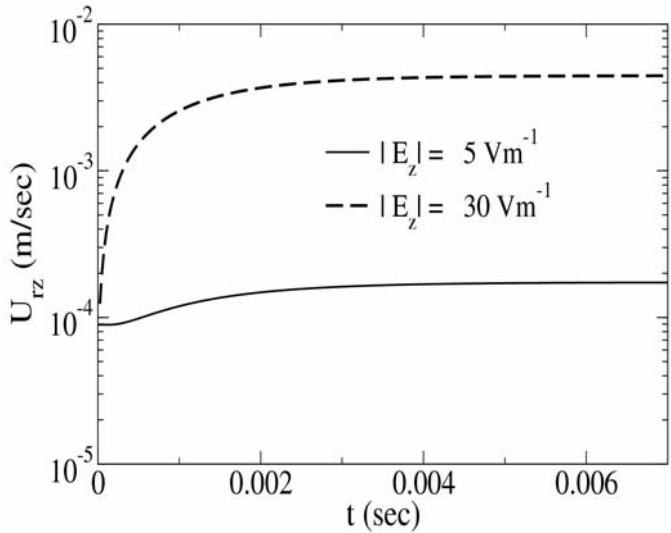

**Figure 10:** Time history of the volume-averaged meridional velocity for different axial electric field intensities $E_z$ in the liquid phase for $V_s = 10^{-4}$ ms$^{-1}$.